\documentclass[twocolumn,showpacs,aps,prl]{revtex4}
\usepackage{graphicx}
\usepackage{amsmath}
\usepackage{amssymb}
\usepackage{epsfig}
\usepackage{multirow}
\usepackage{dcolumn}
\usepackage{bm}
\usepackage{subfigure}

\def\<{\langle}
\def\>{\rangle}

\begin{document}
\title{Non-Fermi liquid behavior in  transport across carbon nanotube quantum dots}

\author{Leonhard Mayrhofer and Milena Grifoni}
\affiliation{Theoretische Physik, Universit\"at Regensburg, 93040 Germany}

\date{\today}

\begin{abstract}
A low energy-theory for non-linear transport in finite-size
single-wall carbon nanotubes, based on a  microscopic model for
the interacting $p_z$ electrons  and successive bosonization, is
presented. Due to the multiple degeneracy of the energy spectrum
diagonal as well as off-diagonal (coherences) elements of the
reduced density matrix contribute to the nonlinear transport. A
four-electron periodicity with a characteristic ratio between
adjacent peaks, as well as nonlinear transport features, in
quantitative agreement with recent experiments, are predicted.
\end{abstract}

 \maketitle


Since their recent discovery  single-wall carbon nanotubes
(SWNTs), cf. e.g. \cite{Saito1998}, have attracted a lot of
experimental and theoretical attention.
In particular, as suggested in the seminal works
\cite{Egger1997,Kane1997},  due to the peculiar one-dimensional
character of their electronic bands,  metallic SWNTs are expected
to exhibit Luttinger liquid behavior at low energies, reflected in
power-law dependence of various quantities
 and spin-charge separation.
Later experimental observations have provided a
confirmation of the theory \cite{Bockrath1999,Postma2001}. As typical of interacting
electron systems in reduced dimension, SWNTs weakly coupled to
leads  exhibit Coulomb blockade  at low temperatures
\cite{Tans1997} with characteristic even-odd 
\cite{Cobden2002} or four-fold periodicity
\cite{Liang2002,Sapmaz2005,Moriyama2005}. In \cite{Sapmaz2005} not
only the ground state, but also several excited states could be
seen in stability diagrams of  closed SWNT quantum dots. Such
two-fold and four-fold character can be qualitatively understood
from symmetry arguments related to the two-fold band degeneracy of
SWNTs and the inclusion of the spin degree of freedom. So far, a
quantitative description has relied on density functional theory
calculations \cite{Ke2003}  or on
 a mean field description of the Coulomb blockade \cite{Oreg2000}.
 In particular, the {position} of the spectral lines in the
stability diagram measured in \cite{Sapmaz2005} was found to be in
quantitative agreement with the predictions in
\cite{Oreg2000}. However, a mean field description may  be {\em  not}
justified for one-dimensional systems. For example, to describe 
the spectral lines of the sample with four-fold periodicity
(sample C) in \cite{Sapmaz2005},  a quite
peculiar choice of the mean field parameters was made, and  a
quantum dot length three times shorter than the measured SWNT length was assumed.
Moreover, to date no quantitative calculation of the nonlinear
current across a SWNT dot has been provided.

In this Letter we investigate spectral as well as { dynamical}
properties of electrons in metallic SWNT quantum dots at low
energies. We start from a microscopic description of metallic
SWNTs
and include Coulomb interaction effects, {\em beyond} mean-field,
by using bosonization techniques \cite{Egger1997,Kane1997},
yielding the spectrum and eigenfunctions of the isolated finite
length SWNT.
 Due to
the many-fold degeneracies of the  spectrum, the current-voltage
characteristics is obtained
 by solving  equations of motion for the  reduced density
 matrix (RDM)
 including off-diagonal elements. Analytical results for the 
 conductance are provided, which account for
  the different heights of the conductance peaks in
\cite{Sapmaz2005}.
Moreover, we can quantitatively reproduce all the spectral lines
seen in sample C in \cite{Sapmaz2005} by solely using the two
ground state addition energies provided in that work. The derived
level spacing is in agreement with the measured SWNT length.

To start with, we consider the total Hamiltonian
\begin{equation}
 H=H_{\odot}+H_{s}+H_{d}+H_{T}+H_{\rm gate},
 \end{equation}
 where
$H_{\odot}$ is the interacting SWNT Hamiltonian (cf. Eq.
(\ref{eq:H SWNT}) below) and $H_{s/d}$ describe the isolated
metallic source and drain contacts as a thermal reservoir of
non-interacting quasi-particles. Upon absorbing terms proportional
to external source and drain voltages $V_{s/d}$, they read
($l=s,d$)
$H_{l}=\sum_{\sigma\vec{q}}\varepsilon_{\vec{q},l}c_{\vec{q}\sigma
l}^{\dagger}c_{\vec{q}\sigma l}$,
 where $c_{\vec{q}\sigma l}^{\dagger}$ creates a quasi-particle
with spin $\sigma$ and energy
$\varepsilon_{\vec{q},l}=\varepsilon_{\vec{q}}-eV_{s/d}$ in
lead $s/d$. The transfer of electrons between the leads and the
SWNT is taken into account by
\begin{equation}
H_{T}=\sum_{l=s,d}\sum_{\sigma}\int\textrm{d}^{3}r\left(T_{l}(\vec{r})
\Psi_{\sigma}^{\dagger}(\vec{r})\Phi_{\sigma
l}(\vec{r})+\textrm{h.c.}\right), \label{eq:H_T}
\end{equation}
where $\Psi_{\sigma}^{\dagger}$ and $\Phi_{\sigma
l}^{\dagger}(\vec{r})=\sum_{\vec q} \phi_{\vec q}^{*}(\vec{r})c_{\vec{q}\sigma
l}^{\dagger}$ are electron creation operators in the SWNT and in
lead $l$, respectively, and $T_{l}(\vec{r})$ describes  the
transparency of the tunneling contact $l$.
Finally, $H_{\rm gate}=-\mu_g{\cal N}_c$ accounts for a  gate
voltage capacitively coupled to the SWNT, with ${\cal N}_c$
counting the total electron number
in the SWNT.\\
{\em SWNT Hamiltonian}. In the following the focus is on armchair
SWNTs at
 low energies. Then, if periodic boundary conditions are applied,  only the gapless energy
sub-bands nearby the Fermi points $F=\pm {\vec K_0}=\pm K_0 \hat e_x$
with ${\hat e_x}$ along the
nanotube axis, are relevant \cite{Egger1997,Kane1997}.
To each Fermi
point two different branches $r=R/L$ are associated
 to the Bloch waves
 $\varphi^{}_{R/L, F,
\kappa } (\vec{r})=e^{i\kappa x}\varphi_{R/L,  F} (\vec{r})$,
where $\kappa$ measures the distance from the Fermi points $\pm K_0$
\cite{Egger1997} 
(Fig. 1a left). In this Letter, however, we are
  interested in finite size effects. Generalizing \cite{Fabrizio1995} to the case of
SWNTs we introduce  standing waves which fulfill {\em open}
boundary conditions 
 (Fig. 1a right):
\begin{equation}
\varphi^{OBC}_{{\tilde R/\tilde L},\kappa}
(\vec{r})=\frac{1}{\sqrt 2} \left[ \varphi^{}_{R/L, K_0, \kappa
}(\vec{r}) - \varphi^{}_{L/R, -K_0, -\kappa} (\vec{r}) \right]\;,
\end{equation}
with quantization condition
 $\kappa = \pi (m_\kappa+\Delta)/L$, $m_\kappa$ an
integer, and $L$ the SWNT length. The offset parameter $\Delta$
occurs  if $K_0 \neq \pi n /L$,  and is responsible for the energy
mismatch between the $\tilde R$ and $\tilde L$ branches. Including
the spin degree of freedom, the electron operator reads
 \begin{equation}
\Psi(\vec{r})=\sum_{\tilde{r}=\tilde{R},\tilde{L}}\sum_{\kappa,\sigma}\varphi_{\tilde{r}\kappa}^{OBC}
(\vec{r})c_{\tilde{r}\sigma\kappa}=:\sum_{\sigma}\Psi_{\sigma}(\vec{r})\label{elektron
operator def}\;,
\end{equation}
 with $c_{\tilde{r}\sigma\kappa}$  the operator which annihilates
$\left|\varphi_{\tilde{r}\kappa}^{OBC}\right\rangle
\left|\sigma\right\rangle $. The interacting SWNT Hamiltonian then
reads
\begin{eqnarray}
H_{\odot} & = & \hbar
v_{F}\sum_{\tilde{r}\sigma}\textrm{sgn}(\tilde{r})
\sum_{\kappa}\kappa c_{\tilde{r}\kappa\sigma}^{\dagger}c_{\tilde{r}\kappa\sigma}+\label{eq:H SWNT} \\
 \!\!\!&& \!\!\!\!\!\!\!\!\!\!\!\!\!\!\frac{1}{2} \sum_{\sigma\sigma'}\int \!d^{3}r\!\int \!d^{3}r'
\Psi_{\sigma}^{\dagger}(\vec{r})
\Psi_{\sigma'}^{\dagger}(\vec{r}')V(\vec{r}-\vec{r}')\Psi_{\sigma'}(\vec{r}')\Psi_{\sigma}(\vec{r})\;,
\nonumber
\end{eqnarray}
 with $v_F$  the Fermi velocity. We
introduce
 1D operators
\[
\psi_{\tilde{r}F\sigma}(x)=\frac{1}{\sqrt{2L}}\sum_{\kappa}e^{i\textrm{sgn}(F)\kappa
x}c_{\tilde{r}\sigma\kappa},\]
 in terms of which the electron operator in (\ref{elektron operator def}) becomes
 \begin{equation}
\Psi_{\sigma}(\vec{r})=\sum_{F,\tilde{r}=\pm}\textrm{sgn}(F)\sqrt{L}{\varphi}_{\textrm{sgn}(F)\tilde
r , F}(\vec{r})\psi_{\tilde{r}F\sigma}(x), \label{psi sigma in
terms of1D}
\end{equation}
where we used the convention that $R/L = \pm 1$,
$\tilde{R}/\tilde{L }=\pm 1$. Upon inserting (\ref{psi sigma in
terms of1D}) into (\ref{eq:H SWNT}), integration   over the
coordinates perpendicular to the tube axis yields the interacting
Hamiltonian expressed in terms of 1D operators and an effective 1D
interaction $V_{eff}(x,x')$. Using standard bosonization
techniques \cite{Egger1997,Kane1997}
$H_{\odot}$ can now be diagonalized
 when keeping
  only forward scattering processes associated to
$V_{eff}(x,x')$.
\begin{figure}[htbp]
\epsfxsize=8cm
\epsfysize=7cm
\center
\includegraphics[width=0.47\textwidth,bb=0 20 321 148]{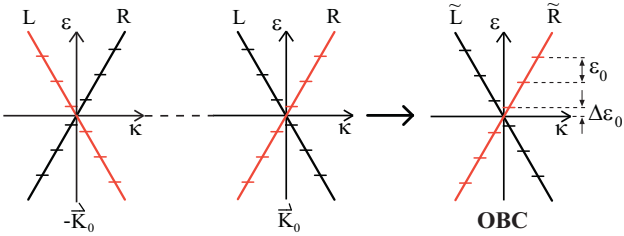}
\caption{Energy spectrum of a SWNT with open boundary
conditions (right) described in terms of left ($\tilde L$)
and right ($\tilde R$) branches. It is constructed from  suitable
combinations of travelling waves whose spectrum is shown on the
left side.
\label{f1}}
\end{figure}
It reads
\begin{eqnarray}
H_{\odot} & = & \frac{1}{2}E_c\mathcal{N}_{c}^{2}+\varepsilon_0\sum_{\tilde{r}\sigma}
\left(\frac{\mathcal{N}_{\tilde{r}\sigma}^{2}}{2}+\Delta\textrm{sgn}(\tilde{r})
\mathcal{N}_{\tilde{r}\sigma}\right)\nonumber \\
 & + & \sum_{q>0}\sum_{j=c,s}\sum_{\delta=\pm}\varepsilon_{j\delta q}a_{j\delta q}^{\dagger}a_{j\delta q}
  \;,\label{eq:Hac-diagonal}\end{eqnarray}
where the first line is the fermionic contribution
and represents the energy cost, due to
Pauli's principle and the Coulomb interaction, of adding new
electrons to the system.
  Specifically,  $
\mathcal{N}_{\tilde{r}\sigma}=\sum_{\kappa}c_{\tilde{r}\sigma\kappa}^{\dagger}c_{\tilde{r}\sigma\kappa},$
is the operator  that counts the number of electrons in the
$(\tilde{r}\sigma)$-branch,  $ \mathcal{N}_c=\sum_{\tilde{r}
\sigma}
 \mathcal{N}_{\tilde{r}\sigma}$  yields the total electron number, and
  $\varepsilon_0=\hbar
v_{F}\frac{\pi}{L}$ is the free-particle level spacing (see Fig. 1).
 The term $E_c=W_{00}$
is the SWNT charging energy, where
 $W_{qq}=\frac{1}{L^2}\int_0^Ldx\int_0^L dx'
 V_{eff}(x,x')\cos(qx)\cos(qx')$.
 The second line of (\ref{eq:Hac-diagonal}) describes bosonic excitations in terms
 of the bosonic operators $a_{j\delta q}$.  Four
channels
are associated to total  ($j\delta=c+$, $s+$) and relative
$(j\delta= c-,s-)$ (with respect to the occupation of the $\tilde R$
and $\tilde L$ branch) charge and spin excitations. Generalized
spin-charge separation occurs, since for three of the channels the
energy dispersion is the same as for the noninteracting system, $
\varepsilon_{j\delta q}=\hbar v_{F} q =\hbar
v_{F}\frac{\pi}{L}n_{q}=:\varepsilon_{0}n_{q}$, ($n_{q}$ a
positive integer),
 but the $(c+)$ channel  is affected by the interaction with
$\varepsilon_{c+\,
q}=\varepsilon_{0}n_{q}(1+8W_{qq}/\varepsilon_{0})^{1/2}$.
  The eigenstates
are \begin{equation}
\prod_{q>0,j\delta}\left(m_{j\delta q}!\right)^{-1/2}\left(a{}_{j\delta
q}^{\dagger}\right)^{m_{j\delta q}}
\left|\vec{N},\vec{0}\right\rangle
=:\left|\vec{N},\vec{m}\right\rangle
,\label{eq:eigenstates}\end{equation} where
$|\vec{N},\vec{0}\rangle $ has no  bosonic
excitations and $\vec{N}$ defines the number of electrons in
each of the branches $(\tilde{r}\sigma).$

{\em Dynamics.} Our starting point to describe transport in SWNTs
is the exact equation of motion
\begin{equation}
i\hbar\frac{\partial\rho^{I}(t)}{\partial
t}=Tr_{leads}\left[H_{T}^{I}(t),W^{I}(t)\right],\label{eq:liouville-diff}\end{equation}
for the reduced density matrix (RDM) $
\rho_{}^{I}=Tr_{leads}W^{I}$ of the SWNT. Here $W^{I}(t)$ is the
density matrix of the whole system consisting of the leads and the
quantum dot, and $Tr_{leads}$ indicates the trace over the lead
degrees of freedom.
 The apex $I$ denotes the interaction
representation with $H_{T}$ from (\ref{eq:H_T}) as the
perturbation.
 We make  the following
  approximations: i) We
 assume
weak coupling to the leads, and treat
$H_{T}$  up to second order, i.e., we consider the leads as
reservoirs which stay in thermal equilibrium and make the
factorization ansatz
$W^{I}(t)=\rho_{}^{I}(t)\rho_{s}\rho_{d}=:\rho_{}^{I}(t)\rho_{leads}$
where
$\rho_{s/d}= Z_{s/d}^{-1}
e^{-\beta(H_{s/d}-\mu_{s/d}\mathcal{N}_{s/d})}$,
with $Z_{s/d}$ the partition function and $\beta$ the inverse
temperature.
 ii) Being
interested in long time properties, we can make the so called
Markov approximation, where the time evolution of
$\dot{\rho}^{I}(t)$ is only local in time.
%
iii) Since we know the eigenstates $|\vec{N},\vec{m}\rangle$ of
$H_{\odot}$, it is convenient to calculate the time evolution of
$\rho_{}^I $ in this basis. We assume that matrix elements between
states representing a different number of electrons (charge
states) in the SWNT and with different energies  vanish.
Coherences between degenerate states with the same energy $E$ are
retained!
Hence we can divide $\rho_{}^{I}(t)$ into block matrices
$\rho_{nm}^{I, E_N}(t),$ where $E,N$ are  the energy and number of
particles in the degenerate eigenstates $\left|n\right\rangle$,
$\left|m\right\rangle$.
We arrive at equations of  the  Bloch-Redfield form
\begin{multline}
\dot{\rho}_{nm}^{I,E_N}(t) = - \sum_{kk'}R_{nm\, kk'}^{E_N}\
\rho_{kk'}^{I,E_N}(t)\\
+ \sum_{E'}\sum_{M=N\pm 1}\sum_{kk'} R_{nm\, kk'}^{E_N\,
E'_M}\rho_{kk'}^{I,E'_M}(t),\label{eq:generalized_Mequ}\end{multline}
where $k,k'$ run over all degenerate states with fixed particle
number. The Redfield tensors are given by $(l=s,d)$
 \begin{equation}
R_{nm\, kk'}^{E_N}  =
\sum_{l}\sum_{E',M,j}\left(\delta_{mk'}\Gamma_{l, njjk}^{(+)E_N\,
E'_M}+ \delta_{nk}\Gamma_{l , k'jjm}^{(-)E_N\, E'_M}\right)
\label{eq:Blochredf_1},
\end{equation} and
$ R_{nm\, kk'}^{E_N\, E'_M}=\sum_{l,\alpha=\pm} \Gamma_{l ,
k'mnk}^{(\alpha) E'_M\, E_N} $, where  the quantities $\Gamma_{l,
njjk}^{(\alpha)E_N\, E'_M}$ are transition rates from a state with
 $N$ to a state  with $M$ particles.
Known the stationary density matrix $\rho_{st}^{I}$, the current (through lead
$l$) follows from
\begin{equation}
I= 2 {\rm Re}\sum_{N,E,E'}\sum_{n k j}\left(\Gamma_{l,
njjk}^{(+)E_N\, E'_{N+1}}-\Gamma_{l, njjk}^{(+)E_N\,
E'_{N-1}}\right)\rho_{kn,st}^{I,E_N}.\label{eq:current}
\end{equation}
iv) We exploit the localized character of the transparencies
$T_l(\vec{r})$ in Eq. (\ref{eq:H_T}), and make use of the slowly
varying nature of the operator $\psi_{\tilde r F \sigma}(x)$ in
Eq. (\ref{psi sigma in terms of1D}).
This enables us to evaluate the 1D operator at the SWNT contacts
and pull it out from the space integrals which enter the
definition of the transition rates. It holds   $ \langle r|
\psi_{\tilde r \sigma F}(x=0)|s \rangle  := \left(\psi_{\tilde r
\sigma}\right)_{rs}^{E_{N} E'_{N+1}}$;  $\langle r| \psi_{\tilde r
\sigma F}(x=L)|s \rangle  = e^{-i\pi{\rm sgn} (F)\{ {\cal
N}_{\tilde r \sigma}{\rm sgn}(\tilde r) +
\Delta\}}\left(\psi_{\tilde r \sigma}\right)_{rs}^{E_{N} E'_{N+1}}
$ for the matrix elements
between the
states $\left|r\right\rangle$, $\left|s\right\rangle$ with energy
$E$, $E'$ and particle number $N$, $N+1$, respectively.  We thus can introduce 
\begin{eqnarray}
\lefteqn{\Phi_{l\tilde r \tilde r'}(\varepsilon)=  \int d^{3}r\int
d^{3}r' {T}_l(\vec{r}) T_l(\vec{r'})
\sum_{\vec{q}|_{\varepsilon}}\phi_{l\vec{q}}(\vec{r})\phi_{l\vec{q}}^{*}(\vec{r'})}\nonumber
\\ && \times \sum_{FF'}{\rm sgn}(FF')  \varphi_{{\rm sgn}(F)\tilde
r,F}(\vec{r})\varphi_{{\rm sgn}(F')\tilde
r',F'}(\vec{r'})\eta_l(\Delta),\nonumber
\end{eqnarray}
 to describe the
influence of the geometry of a tunneling contact at the tube end.
The term $\eta_l(\Delta)=e^{i\pi {{\rm sgn}(F-F')\Delta(1-\delta_{l,d })}}$ accounts for the mismatch
$\Delta$.
  Assuming a 3D electron gas in the leads,
e.g. of gold, we find that for a realistic range of energies is
$\Phi_{l \tilde r \tilde r'}(\varepsilon)=\delta_{\tilde r \tilde
r'}\Phi_l$, i.e. the leads are "unpolarized".  We thus obtain
\begin{multline}
\Gamma_{l, rss'r'}^{(\pm) E_N\, E'_{N+1}}
=\frac{1}{\hbar^{2}}\sum_{\tilde r \sigma} \int d
\varepsilon \rho^\oplus_l(\varepsilon)\Phi_{l}(\varepsilon)
\left(\psi_{\tilde r  \sigma }\right)_{rs}^{E_N\, E'_{N+1}}  \\
\times\left(\psi_{\tilde r  \sigma}^{\dagger}\right)_{s'r'}^{
E'_{N+1}\, E_N}
\int_{0}^{\infty}dt'e^{\pm\frac{i}{\hbar}\left(\varepsilon-eV_{l}-\left(E'-E\right)\right)t'}\;,
\label{eq:gamma_lEnEnp1}\end{multline} with $\rho^\oplus_l(\varepsilon)=\rho_l(\varepsilon)f(\varepsilon),$ where $\rho_l(\varepsilon)$
is the density of energy levels in lead $l$, and $f(\varepsilon)$
the Fermi function. Alike,
\begin{multline}
\Gamma_{l, rss'r'}^{(\pm) E_N\, E'_{N-1}}
=\frac{1}{\hbar^{2}}\sum_{ \tilde r\sigma} \int d
\varepsilon
\rho^\ominus_l(\varepsilon)\Phi_{l}(\varepsilon)
\left(\psi^\dagger_{\tilde r  \sigma}\right)_{rs}^{E_N\, E'_{N-1}}  \\
\times\left(\psi_{\tilde r \sigma} \right)_{s'r'}^{E'_{N-1}\, E_N}
\int_{0}^{\infty}dt'e^{\mp\frac{i}{\hbar}\left(\varepsilon-eV_{l}+\left(E'-E\right)\right)t'}\,,
\label{eq:gamma_lEnEnp2}\end{multline} with $\rho^\ominus_l(\varepsilon)=\rho_l(\varepsilon)\left(1-f(\varepsilon)\right).$

 {\em When are coherences needed?} Eqs.
 (\ref{eq:generalized_Mequ}) with (\ref{eq:current})  show
 that coherences (in the energy basis)   enter the
 evaluation of the current. In the  low bias and temperature
 regime $k_BT, eV:=e(V_s-V_d) \ll \varepsilon_0$, however, where
 only ground states contribute to the current, because of 
$\langle \vec{N},\vec{0}\;| \psi^\dagger_{\tilde {r}
\sigma}\psi_{\tilde {r}  \sigma} |\,\vec{N'},\vec{0}\rangle =
(1/2L)\delta_{\vec{N'},\vec{N}}$, only diagonal elements of the
RDM contribute.
 Hence, due to the "unpolarized" character of the
 leads,  the commonly used master equation
 (CME) with population's dynamics only is valid. At larger
 biases   coherences should be included \cite{footnote}. 
 
In the following we focus on the case $\Delta\approx 0$, relevant
to explain the experimental results for sample C in
\cite{Sapmaz2005}.

 {\em Low bias regime (CME is valid)}.  At low bias
  the current can be obtained by looking to transitions between ground states
  with
 $N$ and $N+1$ particles and energies $E_N^0, E_{N+1}^0$.
  Then, the matrix element $(\psi_{\tilde r
\sigma})_{\vec{N'}\vec{N}}$ 
is non zero  only
if $\vec{N'}=\vec{N}- \hat{e}_{r\sigma}$ , with $
\hat{e}_{r\sigma}$  the unit vector, and 
\begin{equation}
\sum_{\vec{N'}}\sum_{\tilde r \sigma}(\psi^\dagger_{\tilde r
\sigma})_{\vec{N}\vec{N'}}(\psi_{\tilde r
\sigma})_{\vec{N'}\vec{N}} =\frac{1}{2L}C_{N,N'}\;.
\end{equation}Here   $C_{N,N'}$ is the number of ground states with  $N'$
particles whose configurations $\vec{N'}$ differ from the
fermionic configuration of a given ground state with $N$ electrons
only by a unit vector.
With $N=4m,4m+1,4m+2,4m+3$ one finds $C_{N,N+1}=4,3,2,1$ and
$C_{N+1,N}=1,2,3,4$, respectively.  We also notice that all ground
states with $N$ particles are populated with equal probability,
such that we can introduce the occupation probability $P_N(t)=
d_N\rho_{\vec{N},\vec{N}}(t)$, where $d_N$ is the degeneracy of
the ground states with particle number $N$. It holds
$d_N/d_{N+1}=C_{N+1,N}/C_{N,N+1}$. The corresponding CME for
$P_N(t)$ can now be easily solved and the current evaluated in
analytic form. We find
\begin{equation}
 \left|I_{N,N+1}\right| =\frac{e \Delta f\,C_{N,N+1}C_{N+1,N}\gamma_{s}\gamma_{d}}{\sum_{l=s,d}\gamma_l[C_{N,N+1}f(\varepsilon_l)
 +C_{N+1,N}(1-f(\varepsilon_l))]}, 
 \label{eq:INNp1exact}\end{equation} where
 $\Delta f=
\left|f_{}\left(\varepsilon_s \right)-f_{}\left(\varepsilon_d
\right)\right|$, $\varepsilon_l=eV_l-\Delta E$, and $\Delta E=
E_{N}^{0}-E_{N+1}^{0}$. Moreover,
$\gamma_l=(\pi/L\hbar)\Phi_l\rho_l(0)$. This expression can be
further simplified in the  regime $\left|eV\right|\ll
kT\ll\varepsilon_{0}$ where the linear conductance $G_{N,N+1}$ is
 obtained by linearizing $\Delta f$ in $V$, and by evaluating
the remaining quantities in (\ref{eq:INNp1exact}) at zero bias.
%
The conductance trace  exhibits four-electron periodicity (Fig.
2a), with two equal in height central peaks for the transitions
$N=4m+1 \to N+1$,   $N=4m+2\to N+1$, and  two smaller
peaks for  $N=4m\to N+1$,  $N=4m +3 \to N+
1$ also equal in height. The relative height between central and
outer peaks is $ G_{4m+1,\
4m+2}^{max}/G_{4m,4m+1}^{max}=27/(10+4\sqrt{6})\approx1.36$,
independent of the ratio $\gamma_s/\gamma_d$.

   In the bias regime
$\varepsilon_{0}\gg\left|eV_{l}\pm\Delta E\right|\gg kT$ is $\Delta
f =1$. If e.g. $eV_{s}-\Delta E<0$ and $eV_{d}-\Delta E>0,$
such that tunneling is preferable rom
 source to drain, we find
$I_{N,N+1}=e C_{N,N+1}C_{N+1,N}\gamma_{s}\gamma_{d}/
(\gamma_{s}C_{N,N+1}+\gamma_{d}C_{N+1,N}).$
In this regime, the nonlinear conductance will still
exhibit four-electron periodicity, Fig. 2b.
For $\gamma_s=\gamma_d$ one still expects two equal central peaks and
two smaller outer peaks with ratio $3/2$. If $\gamma_s\neq
\gamma_d$ this latter symmetry is lost.
If we invert the sign of the bias voltage, the current is obtained by exchanging $\gamma_s$ with
$\gamma_d$.

\begin{figure}[htbp]
\center
\includegraphics[%
  width=0.47\textwidth,
  keepaspectratio]{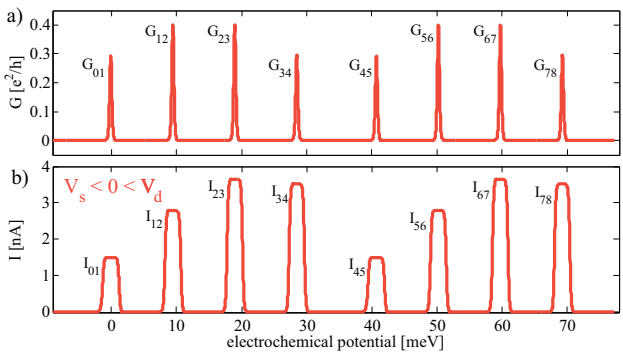}
\caption{a) Conductance vs. electrochemical potential in the linear regime
$eV\ll k_BT\ll \varepsilon_0$. Despite asymmetric contacts, the two central peaks and the two outer
peaks have equal height. b) Current in the regime
$\varepsilon_{0}\gg\left|eV_{l}\right|\gg kT$. Asymmetry effects
become visible. Four-electron periodicity is still observed. 
Parameters are $E_c=9.5$ meV; $k_BT= 0.10$ meV,
$\varepsilon_0=2.9$ meV, $\gamma_s=5\gamma_d=4.9\cdot 10^{10} s^{-1}$.  \label{f2}}
\end{figure}
{\em High bias regime.} In the bias regime $eV > \varepsilon_0$
states with bosonic as well as fermionic excitations contribute to
transport. An analytical treatment is not possible, except to
define the position of the various excitation lines. The resonance
condition for tunneling in/out of lead $\l$ is as usually given by
$ eV_{l}+\Delta E^{\pm}=0$, where $\Delta
E^{\pm}=\pm\left(E_{N\pm1}-E_{N}\right).$ 
Besides the resonance condition, also the overlap integral between
initial and final state determines the rates, and hence the
"active"  resonance lines contributing to the current. 
Fig 3a shows the current in a bias voltage-electrochemical potential plane for 
 the symmetric case $\gamma_s=\gamma_d$.
By choosing the addition energies provided in \cite{Sapmaz2005} $
\Delta\mu_1=E_c=9.5$ meV and $\Delta \mu_2=13.4$ meV, we can
reproduce all the excitation lines from sample C in
\cite{Sapmaz2005}. Moreover, we find a level spacing
$\varepsilon_0=2.9$ meV, which well agrees with the estimated
length for sample C of $750$ nm. We compare with the mean field
parameters: to fit the data, an unusually large exchange
interaction $J=2.9$ meV as well as a band shift
$\varepsilon_0\Delta\approx J$ had to be assumed in
\cite{Sapmaz2005} (in our theory is $\varepsilon_0\Delta\approx J
\approx 0)$. This yields a level spacing three times larger than
the one obtained from our treatment and not consistent with the
measured SWNT length.

Finally, the effect of the coherences induced by the bosonic
excitations is shown in Fig. 3b, where a difference plot for the
current with and without coherences is shown. Though the
coherences do not qualitatively change the current, they do have a
quantitative influence in a region of intermediate bias $V$. A
further indication for non-Fermi liquid behavior could lie in
negative differential  (NDC) features originating from spin-charge
separation, as predicted for a spinful Luttinger liquid quantum
dot  \cite{Cavaliere2004}. Asymmetric contacts are a necessary
requirement.
We checked these predictions as possible explanation of the NDC
seen in \cite{Sapmaz2005}. We confirm that (also for non-relaxed
bosons) NDC occurs. However, very large asymmetries must be
assumed. Moreover, in contrast to the experiments, all the NDC
lines have the same slope.

To conclude, we discussed linear and nonlinear transport in SWNT
quantum dots using a bosonization approach. Our results are in
quantitative agreement with experimental findings in
\cite{Sapmaz2005}. Further work to explain the nature of the NDC
seen in \cite{Sapmaz2005} is needed.
\vspace{-2 pt}
\begin{figure}[htbp]
\center
\center
\includegraphics[%
  width=0.47\textwidth,
  keepaspectratio]{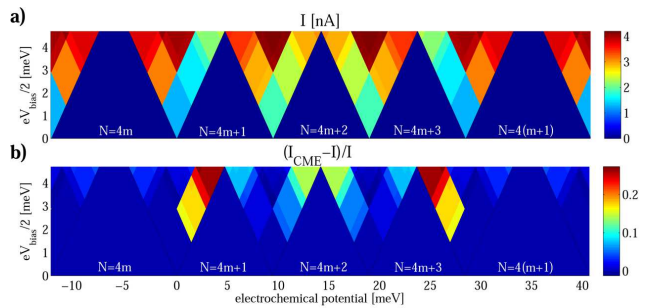}

\caption{a) Current in a bias voltage - electrochemical potential plane for the 
symmetric contacts case. 
 b) Difference plot of the current with and
without coherences. Here $k_BT=0.01$ meV and $\varepsilon_0/\varepsilon_{c+q}\approx 0.21$. Other parameters are as in Fig. 2a. \label{f3}}
\vspace{-12 pt}
\end{figure}

\centerline{***}
 Useful discussions with S. Sapmaz and support by the DFG under the program GRK 638 are acknowledged.


\begin{thebibliography}{12345678}
\bibitem{footnote} In the case of noninteracting electrons and unpolarized leads, the CME is always  correct  if the usual Bloch wave Slater determinants are used as energy eigenstates.
\end{thebibliography}
\end{document}